# Properties of Microelectromagnet Mirrors as Reflectors of Cold Rb Atoms

M. Drndic, G. Zabow, C. S. Lee, J. H. Thywissen, K. S. Johnson, M. Prentiss,

and R. M. Westervelt

*Division of Engineering and Applied Sciences and Department of Physics*

*Harvard University, Cambridge MA 02138*

P. D. Featonby, V. Savalli, L. Cognet, K. Helmerson[a], N. Westbrook, C. I. Westbrook,

W. D. Phillips[a], and A. Aspect

*Laboratoire Charles Fabry, Institut d'Optique Théorique et Appliquée,*

*B.P.147, 91403 Orsay Cedex, France*

Cryogenically cooled microelectromagnet mirrors were used to reflect a cloud of free-falling laser-cooled $^{85}$Rb atoms at normal incidence. The mirrors consisted of microfabricated current-carrying Au wires in a periodic serpentine pattern on a sapphire substrate. The fluorescence from the atomic cloud was imaged after it had bounced off a mirror. The transverse width of the cloud reached a local minimum at an optimal current corresponding to minimum mirror roughness. A distinct increase in roughness was found for mirror configurations with even versus odd number of lines. These observations confirm theoretical predictions.

---

[a] Permanent address National Institute of Standards, Gaithersburg, MD 20899.

In recent years there has been an increased effort to design and construct optical elements for neutral atoms including mirrors, gratings, and lenses [1]. Atom mirrors have been demonstrated using evanescent light fields [2,3] and evanescent static magnetic fields [4-7]; curved mirrors have been formed using evanescent light fields [2] and magnetized floppy discs [6] to focus and gravitationally trap atoms. Exponentially decaying magnetic field mirrors were proposed to reflect neutrons [8], atoms, and molecules [9], and permanent magnet [4-7] and microelectromagnet atom mirrors [10-12] have been demonstrated. Microelectromagnet atom mirrors consisting of parallel lines of alternating current [10-12] can additionally create time-dependent potentials thus forming adaptive atom optics [11]. Ultimately, microelectromagnet mirrors promise an impressive "flatness" for several reasons: 1) microfabrication can produce precisely controlled wire shapes and geometries on substrates with excellent surface smoothness, 2) the detrimental effect of possible wire irregularities on magnetic field equipotentials is suppressed by electrical current conservation [10,11], and 3) the strength of mirrors can be adjusted by applying time-dependent currents.

In this Letter, we describe the properties of microelectromagnet mirrors to reflect a cloud of free-falling laser-cooled Rb atoms at normal incidence. We imaged the reflected cloud, dropped from a magneto-optical trap, and measured its size as a function of mirror current. This is the first experimental study of the influence of the finite size of the mirror on the reflective properties of microelectromagnet mirrors consisting of current-carrying wires. The experiment confirms recent theoretical predictions of surprising and nontrivial dependence of mirror roughness on mirror current and on mirror parity [13]: the measured roughness reaches a local minimum at an optimal current, rather than decreasing monotonically as current increases, and mirrors with an odd $N$ number of wires are smoother than mirrors with even $N$.

Figure 1(a) shows a photograph of a microelectromagnet atom mirror consisting of a serpentine pattern of current-carrying Au wires on a sapphire substrate, together with a schematic diagram of the wire pattern (Fig. 1(b)) and a profilometer trace of the wire profile (Fig. 1(c)). The details of fabrication, magnetic field calibration, and thermal properties have been reported earlier [10]. Mirrors have been made from normal metals (Au, Ag, Cu) and superconductors (Nb) with periods [$a$ in Fig. 1(c)] ranging from $a = 12$ µm to $a = 200$ µm, and covering areas up to 1 cm$^2$. We use optical lithography to control precisely the mirror geometry across large surface areas. To achieve the high current carrying capabilities of our mirrors, the wires are made thicker by electroplating, heat sunk onto a sapphire substrate, and cryogenically cooled as described in [10]. The fabricated mirrors used in this experiment thus permitted high current densities up to ~ 10$^7$ A/cm$^2$ and magnetic fields to $B$ ~ 0.1T with gradients $|\nabla B|$ ~ 100 T/cm. The wire widths $w$, heights $h$, and spacings ($a/2-w$) can be continuously varied in fabrication by adjusting the electroplating parameters to grow the wires to desired sizes and aspect ratios [14]. This flexibility is important, because recent simulations showed that optimization and control of wire profiles is useful to make high quality mirrors [13].

Unless otherwise noted, the mirror specifications in the present experiment are: active area 1 x 1 cm$^2$ of the serpentine pattern of Au wires, period $a = 200$ µm, $N = 101$ lines, resistance $R = 40$ Ω at room temperature, and $R = 10$ Ω at 77 K. An additional wire segment with the same profile and spacing, with resistance ~ 0.4 Ω at room temperature was added to the original mirror, as shown in Fig.1(b), to allow increasing the number of wires from odd to even during the experiment. Figure 1(c) shows a profilometer trace of a duplicate mirror with the same process parameters and resistance as the one used in the experiment, with the wire width $w \approx 70$ µm and height $h \approx 7$ µm indicated. To avoid damaging the mirror, we have applied currents only up to 3 A at liquid nitrogen temperatures, which corresponds to fields up to ~ 30 mT at the surface of the wires.

The operation of the atom mirror is based on the Zeeman interaction between an inhomogeneous magnetic field and the atomic magnetic dipole moment [9]. For an infinite number of infinitely long wires, the field decays exponentially $B_{inf} = \alpha I e^{-kz} (1 + \varepsilon e^{-2kz} \cos(2kx) + ...)$, where $k^{-1} = a/2\pi$ is the exponential decay length, $z$ is the perpendicular distance from the mirror surface, $x$ is the distance along the mirror and perpendicular to the wires, $I$ is the current in the wires, and $\alpha$ and $\varepsilon$ are coefficients determined by wire shape and spacing [8-10]. The field magnitude and not the direction controls the reflection of slowly moving atoms in the adiabatic regime when the atomic magnetic moment follows the local field direction and the magnetic quantum number $m_F$ of the atomic substate is a constant of motion [8,9]. Consequently, the gradient force due to the magnetic field is $\boldsymbol{F} = -g_F m_F \mu_B \nabla B$, where $\mu_B$ is the Bohr magneton, and $g_F$ is the Lande g factor. Atoms in weak field seeking states ($m_F g_F > 0$) are reflected elastically from high fields near the mirror. As shown below in Fig. 2, deviations from perfect specular reflection are caused by the small corrugations/roughness in the magnetic field equipotentials that deviate from perfectly flat lines parallel to the mirror surface. For an infinitely large microelectromagnet mirror, the effects of the corrugations decrease monotonically with the increasing mirror current [9].

Several authors have noted theoretical differences between the reflective properties of a realistic mirror with a finite number of wires and the ideal mirror with an infinite number of wires (or analogously, in the case of permanent magnet mirrors, between finite and infinite numbers of magnetic domains) [5,12,13]. It has been pointed out [5] that the performance of finite mirrors can be improved by adding compensating wires to the edges of the mirror along the y-direction (see Fig. 1 (b)). Recent theoretical work [13] has shown that the roughness of a finite mirror operated at constant currents (even one with additional compensating wires) reaches a minimum value at a finite current, in contrast to the infinitely large mirror that always becomes less corrugated with increasing currents. Moreover, the roughness of a finite mirror was also predicted to depend critically on whether the number of lines in the mirror array is even or odd. Both of these predictions can be understood by noting that the field of the finite mirror, $\boldsymbol{B}_{\text{fin}} = \boldsymbol{B}_{\text{inf}} - \boldsymbol{B}_{\text{miss}}$, where $\boldsymbol{B}_{\text{inf}}$ is the vector field of an infinite mirror, and $\boldsymbol{B}_{\text{miss}}$ is the vector field due to wires in the infinite mirror that are missing in the finite mirror. Atoms are reflected from surfaces of constant B-field magnitude, which for an ideal infinite mirror are constant $z$ planes:
$\left|\boldsymbol{B}_{\text{ideal inf}}\right| = \alpha I e^{-kz} \left|\cos(kx)\vec{e}_x + \sin(kx)\vec{e}_z\right| = \alpha I e^{-kz}$. In contrast $\boldsymbol{B}_{\text{fin}}$ includes cross-terms between $\boldsymbol{B}_{\text{inf}}$ and $\boldsymbol{B}_{\text{miss}}$ and $\left|dB_{\text{fin}}/dx\right| > \left|dB_{\text{inf}}/dx\right|$. In the case of a finite mirror with an odd (even) number of wires, the current distributions producing $\boldsymbol{B}_{\text{miss}}$ are roughly parallel (antiparallel) on the opposite

sides of the mirror. From symmetry, this results in a $\boldsymbol{B}_{miss}$ that is small (large) near the center of the mirror, making the odd mirror flatter than the even.

Figures 2(a) and 2(b) show computed contours (equipotentials) of constant magnetic field magnitude $B$ above the centers of two microelectromagnet mirrors with even ($N = 102$) and odd ($N = 101$) numbers of wires; the wires in the array are taken for simplicity to have rectangular cross-sections. The field contours exhibit interesting and nontrivial features at heights $z \sim a$ above the mirror. Because the field from an infinite array decays exponentially, the effect of the end wires in a finite serpentine array is felt at a finite height above the mirror center. The first interesting feature, in contrast to the infinite $N$ case, is that the magnetic field contours for finite $N$ in Fig. 2 best approximate straight lines at a finite height above the mirror. Because the height at which atoms of a certain initial velocity are reflected is determined by mirror current $I$, the roughness is smallest at approximately the current which gives a turning point in the region where the equipotentials are most straight. A second interesting feature is that mirrors with an even number of wires $N$ (Fig. 2(a)) have more corrugated equipotentials than those with odd $N$ (Fig. 2(b)).

The experimental setup used to reflect cold atoms from microelectromagnet mirrors is schematically illustrated in Figure 3(a). A cloud of cold $^{85}$Rb atoms in the $5^2 S_{1/2}$ $F = 3$ electronic ground state is accumulated and cooled in a magneto-optical trap (MOT). The atomic setup was the same as in Ref. [3]. The number of trapped atoms is $\sim 3 \times 10^7$ in a volume $\sim 0.5$ mm$^3$ with root-mean-square atom velocity $v_{rms} \sim 36$ mm/s. The atom cloud is released from the MOT with center at a constant height $H = 19.6$ mm above the mirror, and optically pumped during the fall into the $|F = 3, m_F = 3>$ state. A 1.06 mm wide slit mounted $\cong 3$ mm above the mirror defines the width of the atom cloud which reaches the center of the mirror. The mirror is mounted on a copper cold finger and cooled by continuously flowing liquid nitrogen to lower the mirror resistance and increase the thermal conductivity of the mirror and the sapphire substrate. The temperature of the mirror is estimated to be $\sim 100$ K from the known temperature dependence of the mirror resistance. Rectangular current pulses with amplitudes $I$ and duration 5 ms coinciding with the arrival of the atom cloud were applied to the mirror every 1.5 s. The pulsed mode operation together with cryogenic cooling of the mirror significantly reduced the average heat dissipation and allowed for higher currents and field strengths. To ensure that atoms adiabatically follow the magnetic field direction, i.e. remain in the $m_F = 3$ state, a constant holding field $B_h = 10$ µT was applied along the y-direction parallel to the mirror lines and perpendicular to the mirror field (see Fig. 3(a)).

The width of the atom cloud was measured by a horizontal retroreflected probe laser beam during the free fall after the first bounce, and during the fall back before the second bounce. The probe laser beam was positioned at height $Z = 9.4$ mm above the mirror (see Fig. 3 (a)). The spatial extent of the probe beam was 1.4 mm in the z-direction by 1.0 cm in the y-direction, its power density was 1.1 mW/cm$^2$, and it was detuned 5 MHz from the fluorescence line. The fluorescence was imaged by a charge-coupled device (CCD) camera looking down at the mirror along the z-direction [3]. The stray-light background recorded by the camera was subtracted from the fluorescence signal.

Figures 3(b) and 3(c) show typical fluorescence images of the atom cloud together with fluorescence profiles along the x-direction after the bounce (Fig. 3(b)) and as the atom cloud falls back (Fig. 3(c)) before the second bounce on the mirror. The profiles empirically fit Gaussian distributions $F(x) \propto e^{-(x-x_o)^2/2w_{rms}^2}$. From fits to measured profiles such as those in Fig. 3 we obtain the root-mean-square width $w_{rms}$ of the cloud at height $Z = 9.4$ mm above the mirror. Had the atoms been reflected specularly by an ideally smooth infinite mirror, the intrinsic width $w_o$ would then be determined by the initial momentum and spatial distribution of the atoms in the cloud, the drop height, the slit width, and the effect of the probe laser; Monte-Carlo simulations of a specular reflection give $w_o = 2.54$ mm [15]. The actual measured widths of the atomic cloud $w_{rms}$ are larger than $w_o$ because the actual mirror, finite in size, is not ideally smooth. We have used the fitted width of the atomic cloud to study the reflective mirror properties as a function of current and to compare mirrors with odd and even numbers of wires.

Figure 4 shows the experimental width $w_{rms}$ of the reflected atomic clouds falling back under gravity toward the mirror as a function of mirror current $I$ for mirrors with $N = 101$ and $N = 102$ wires. The number of wires was changed during measurements by energizing the additional wire segment shown in Fig. 1(b). As predicted theoretically, the data of Fig. 4 clearly show three mirror properties: (1) a threshold current $I_{th}$ for atom reflection, (2) an optimal current $I_{opt}$ at which the width of the reflected cloud is minimized, and (3) a pronounced increase in width for the mirror with an even number of wires. (1) Reflection is only observed at mirror currents above the threshold $I_{th} \cong 0.35$ A, as shown in Fig. 4. As the current increases, the magnetic field increases, and the height of the turning point for the atom bounce increases. The threshold occurs when atoms no longer hit the wires or substrate. The minimum magnetic field required to reflect $^{85}$Rb atoms in the $m_F = 3$ ground state dropped from $H = 19.6$ mm is calculated from the incident atomic energy to be $B_{th} \cong 3$ mT at the turning point. This value agrees well with a threshold estimated from the computed magnetic field 2.9 mT for $I_{th}$ at the top of 70 x 7 μm² rectangular wires using simple Biot-Savart calculations. Close to threshold the measured width $w_{rms}$ of the reflected cloud is comparable for odd and even mirrors, consistent with the fact that the magnetic field equipotentials for even and odd mirrors shown in Figs. 2(a) and 2(b) are comparable near the wires. (2) In contrast to the simpler infinite mirror case, the transverse width $w_{rms}$ of the atom cloud is minimized at an optimal mirror current. For even $N = 102$ the optimal current is around $I_{opt} \sim 1$ A and for odd $N = 101$ it is around $I_{opt} \sim 1.5$ A, as shown in Fig. 4. From Fig. 2 we can understand that the mirror will be least corrugated when the current is approximately adjusted to give a turning point in the smoothest magnetic field contour region, and that the mirror with even $N$ will have a lower optimal current than the mirror with odd $N$. The increase in the measured width $w_{rms}$ over the width $w_o$ for perfect specular reflection provides an upper bound to the angular spread: $\Delta\theta_{max} = \left(w_{rms}^2 - w_o^2\right)^{1/2} / 2v_z t$, where $v_z$ is the average vertical velocity and $t$ is the time interval; we estimate $\Delta\theta_{max} \approx 16$ mrad for $N = 101$ at $I \sim I_{opt}$. This upper bound to the angular spread is comparable to recent calculations [13] of the rms angular spread $\Delta\theta_{rms} \sim 10$ mrad for finite sized microelectromagnet mirrors with $N = 101$ wires at optimum current. In previous experiments the reported angular spread was 5 mrad for evanescent wave mirrors [16], 6 mrad for permanent magnet mirrors [6], and 45 mrad for electromagnet mirrors [12]. (3) Figure 4 also compares reflected atom cloud widths at various currents from mirrors with even $N = 102$ and odd $N = 101$ numbers of wires. The only difference between the two mirror configurations came from energizing the additional wire at the end. We observe a clear difference in roughness between the two mirrors, the roughness being lower for odd $N = 101$. From Fig. 2 we can see that odd $N$ mirrors have a larger region of more nearly straight equipotentials than even $N$ mirrors and thus are expected to be smoother.

Reflection of atoms was also observed from a smaller mirror with active area 2 x 2 mm$^2$ of the serpentine pattern, $N = 83$, $a = 48$ μm, $w \approx 20$ μm, and $h \approx 3$ μm. The mirror resistance was $R = 60$ Ω at room temperature and $R = 20$ Ω at ~ 100 K. For this small mirror the measured threshold current $I_{th} \approx 90$ mA is smaller than for the large mirror, due to the stronger fields produced near wires with smaller cross sectional area and spacing [10]. The threshold estimated from the computed magnetic field 2.6 mT for $I_{th}$ at the top of 20 x 3 μm$^2$ rectangular wires using the Biot-Savart law agrees well with the threshold computed from the atom energy and with the threshold observed for the larger mirror. Reflection from this small mirror is difficult to analyze in more detail, because the atom cloud was comparable in size to the mirror.

In conclusion, we have 1) reflected cold atoms from microelectromagnet mirrors with odd and even numbers of wires, 2) demonstrated the existence of finite currents which optimize the reflective mirror properties, and 3) shown that mirrors with an odd number of wires are smoother than mirrors with an even number of wires. Improvements of reflective properties might be possible by adding compensating wires along the edges of mirrors with an odd number of wires to mimic the field of an ideal infinite mirror [5,12,13,17], by tailoring the wire shape and geometry [13,14], and by exploiting time-dependent currents. It is important to note that presently attainable currents are already sufficient to exceed the current required for minimum roughness as shown in Fig. 4, and that larger currents are not necessary. By varying the wire spacing across the width of the mirror, focusing microelectromagnet mirrors could be realized on flat substrates. It is interesting to note that in another area of high magnetic fields, pulsed microcoils have recently achieved fields as high as $B \sim 50$ T [18], suggesting that microelectromagnet mirrors could also be useful for more energetic particles.

The authors thank I. Silvera and M. Topinka. This work was supported by the National Science Foundation (NSF) Grants DMR-9809363, PHY-9312572, and PHY-9732449, the CNRS région Ile de France and the European Union TMR network, contract number ERBFMRXCT960002. J.H.T. acknowledges the support from the F. and J. Hertz Foundation. M.D., W.D.P., and G.Z. thank the Institut d'Optique for its hospitality.

**Figure Captions**

FIG. 1 Serpentine microelectromagnet Au mirror on sapphire substrate (a) micrograph, (b) mirror pattern, (c) profilometer scan of a mirror with the same parameters as the one used in the experiment; wire width $w \approx 70$ μm and height $h \approx 7$ μm.

FIG. 2 Computed magnetic field magnitude contours above two microelectromagnet mirrors with (a) even $N = 102$ and (b) odd $N = 101$ wires. The serpentine area is 1 cm$^2$, the period $a = 200$ μm, width $w = 70$ μm, and height $h = 7$ μm. The mirror center is at $x = 0$ and $z$ is the height above the substrate. Subsequent contours are in the ratio $B_{n+1}/B_n = 0.8$, where n = 1 to 20, and $B_1 = I(7.2$ mT/A$)$.

FIG. 3. (a) Experimental setup used to reflect cold Rb atoms from a microelectromagnet mirror; $H = 19.6$ mm and $Z = 9.4$ mm. Fluorescence profiles F along the x-direction and atom cloud images are shown at height $Z$ (b) after the first bounce and (c) during the fall before the second bounce. The solid lines are Gaussian fits.

FIG. 4. Root-mean-square widths $w_{rms}$ of the Rb cloud as it falls after the first bounce vs. mirror current $I$ for mirrors with even ($N = 102$) and odd ($N = 101$) numbers of wires. The computed width $w_o$ for a perfectly specular bounce is the dashed line.

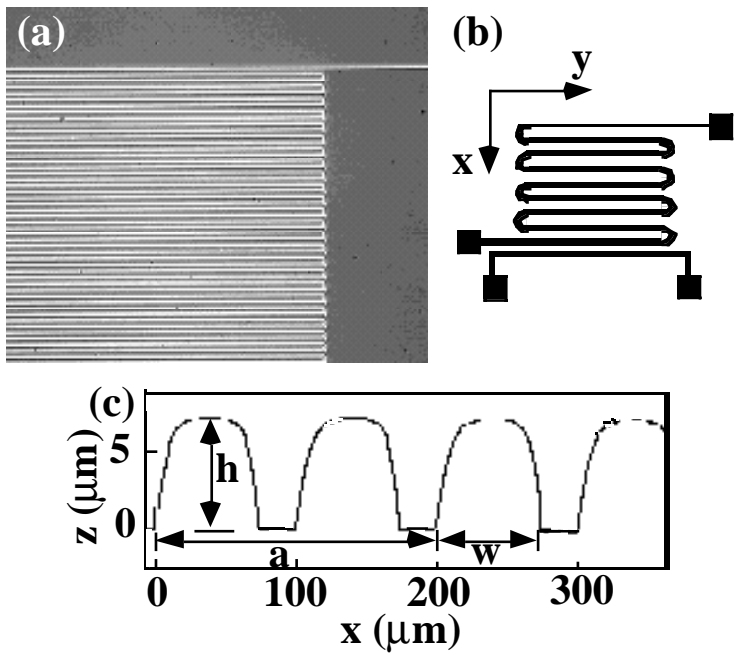

Figure 1
M. Drndic *et al.*
Physical Review A

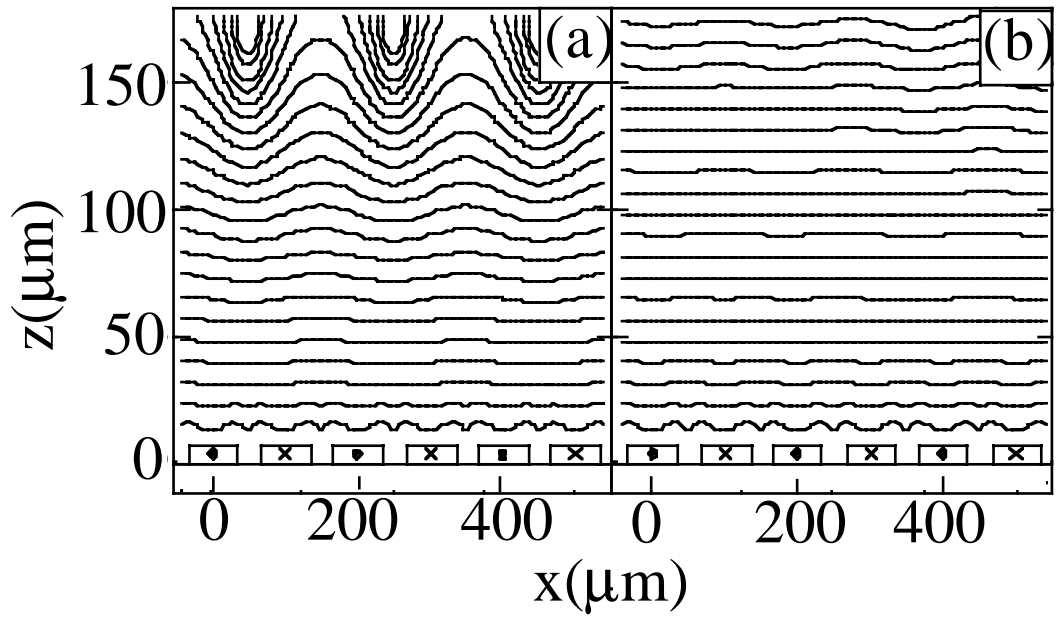

Figure 2
M. Drndic *et al.*
Physical Review A

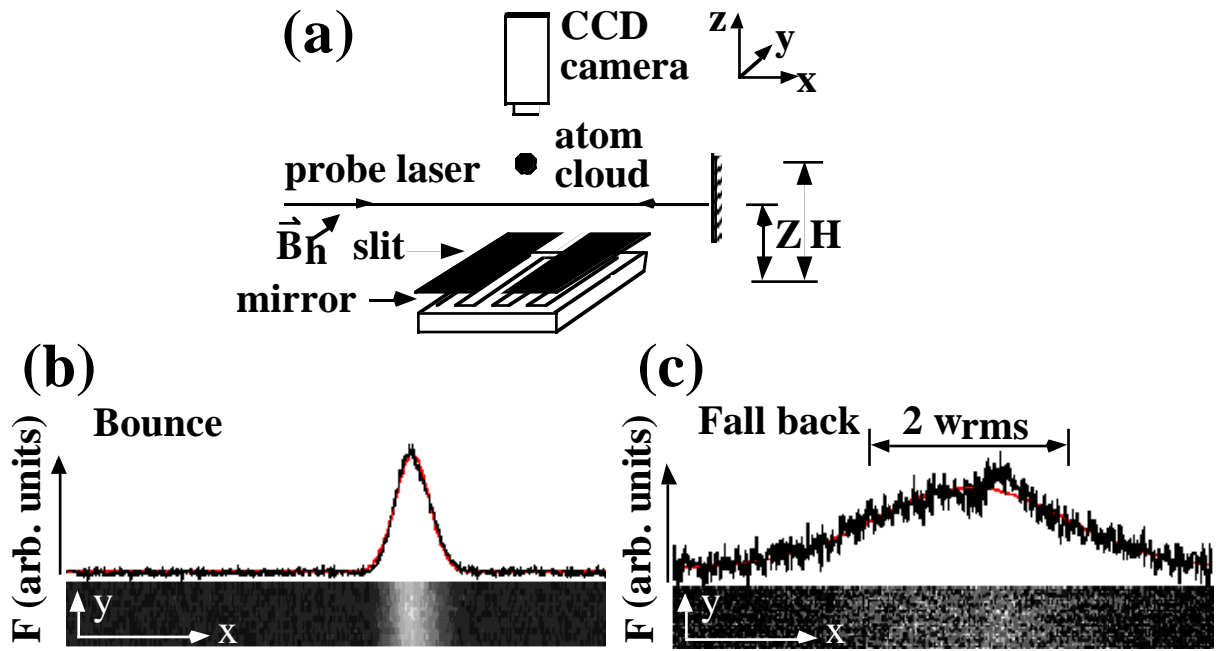

Figure 3
M. Drndic *et al.*
Physical Review A

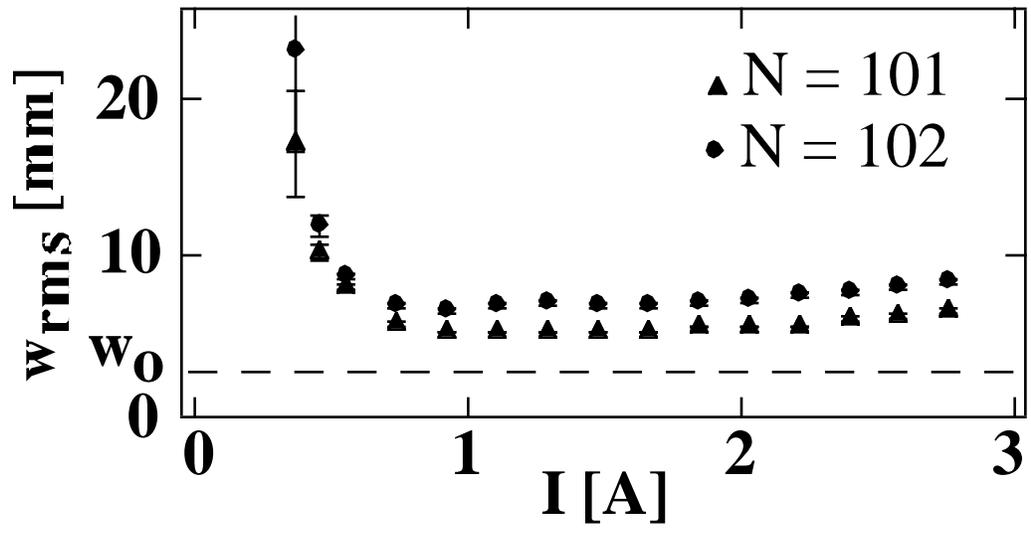

Figure 4
M. Drndic *et al.*
Physical Review A